\documentclass[a4paper,11pt]{article}
\usepackage{bm,amssymb,amsmath,cite,color,verbatim,marvosym,graphicx,multicol}
\usepackage[colorlinks]{hyperref}

\def\lb{\label}
\def\ba{\begin{eqnarray}}
\def\ea{\end{eqnarray}}

\newcommand{\nc}{\newcommand}
\nc{\be}{\begin{equation}} \nc{\ee}{\end{equation}}
\nc{\bea}{\begin{eqnarray}} \nc{\eea}{\end{eqnarray}}
\nc{\disp}{\displaystyle} \nc{\ade}{\mbox{$A$-$D$-$E$}}
\nc{\calN}{{\cal N}} \nc{\calC}{{\cal C}} \nc{\calM}{{\cal M}}
\nc{\calS}{{\cal S}} \nc{\phit}{\hat{\varphi}}
\nc{\chit}{\hat{\chi}} \nc{\hcalN}{\hat{\calN}}
\nc{\hcalS}{\hat{\calS}} \nc{\hS}{\hat{S}}
\nc{\sigmad}{\sigma^\dagger} \nc{\psid}{\psi^\dagger}

\def\non{\nonumber}

\nc{\todo}[1]{\textbf{( #1 )}}

\definecolor{IndianRed}{rgb}{0.8,0.36,0.36}
\def\ir{\color{IndianRed}}
\definecolor{DarkGreen}{rgb}{0,0.5,0}

\newtheorem{defn}{Definition}

\font\tenmsb=msbm10\font\sevenmsb=msbm7 \font\fivemsb=msbm5
\newfam\msbfam
\textfont\msbfam=\tenmsb \scriptfont\msbfam=\sevenmsb
\scriptscriptfont\msbfam=\fivemsb

\def\bra#1{\langle #1|}
\def\ket#1{|#1\rangle}

\renewcommand\eqref[1]{(\ref{#1})}

\DeclareMathOperator{\e}{e}

\def\ii{{\,\rm i}}

\numberwithin{equation}{section}

\begin{document}
\hypersetup{%
    urlcolor=IndianRed, 
    linkcolor=DarkGreen, 
    citecolor=DarkGreen 
} 
\begin{titlepage}
\title{Exact finite size groundstate of the O($n=1$) loop model with open boundaries}

\author{Jan de Gier$^a$\footnote{jdgier@unimelb.edu.au}, Anita Ponsaing$^a$\footnote{a.ponsaing@ms.unimelb.edu.au} and Keiichi Shigechi$^b$\footnote{Keiichi.Shigechi@lmpt.univ-tours.fr}
\bigskip\\
{\small\em
\begin{minipage}{\textwidth}
\begin{itemize}
\addtolength{\itemsep}{-\baselineskip}
\addtolength{\itemsep}{-\parskip}
\item[${}^a$] Department of Mathematics and Statistics, The University of Melbourne, VIC 3010, Australia\\
\item[${}^b$] Laboratoire de Math\'ematiques et Physique Th\'eorique CNRS/UMR 6083, F\'ed\'eration Denis Poisson, Universit\'e de Tours, Parc de Grandmont, 372000 Tours, France
\end{itemize}
\end{minipage}}}

\maketitle

\begin{abstract}
We explicitly describe certain components of the \textit{finite size} groundstate of the inhomogeneous transfer matrix of the O($n=1$) loop model on a strip with non-trivial boundaries on both sides. In addition we compute explicitly the groundstate normalisation which is given as a product of four symplectic characters.
\end{abstract}

\end{titlepage}

{\footnotesize\tableofcontents}
\vfill\newpage

\section{Introduction}

In recent years it has been realised that the groundstate of the transfer matrix of the critical O($n=1$) loop model on the square lattice can be explicitly described for \textit{finite} system sizes \cite{BGN01,RS01}. Such feasibility is rather unusual for critical models,\footnote{Valence bond states (VBS) or dimerized states are common examples for gapped systems. Another well known example is the matrix product stationary state for the asymmetric exclusion process.} and is commonly reserved for free fermion models only. Exact knowledge of the finite size groundstate provides access to the precise size dependence of the most probable configurations of the O($n=1$) model, and hence to finite size correlation functions. Examples of conjectural results for such correlations, based on numerical studies for homogeneous system, can be found in \cite{MitraNGB04}.

Applications of the O($n=1$) model are abundant, the most well known ones being critical bond percolation on the square lattice and the quantum XXZ spin chain at $\Delta=-1/2$, see e.g. \cite{BaxterKW76}. Other closely related models are the stochastic raise and peel model \cite{PearceRGN02} and lattice realisations of models with supersymmetry \cite{Fendley,YangF04,GNPR05} or an underlying logarithmic conformal field theory \cite{RasmusP}. A further interesting connection has been made with the quantum Hall effect \cite{KP}.

Following initial conjectural results in \cite{BGN01,RS01} and \cite{PearceRGN02,RS01b,Gier02,MitraNGB04}, a method was initiated and developed in \cite{DFZJ04,DF05,DFZJ05,Pasq05,ZJ07} for proving a variety of results regarding the O($n=1$) groundstate by generalising to inhomogeneous models. This approach was extended in \cite{DFZJ07,DFZJ07b}, introducing the use of multiple contour integral expressions for certain linear transformations of the groundstate, which has been successful for establishing several conjectures. An alternative approach describing each component of the groundstate in terms of factorisations of non-commuting operators was developed in \cite{GierP07}, see also \cite{KiriL00}.

Although a wide variety of periodic and open boundary conditions has been studied, the case of two open boundaries has so far resisted progress. Here we deal with this case, and will generalise the results for reflecting \cite{DF05} and mixed \cite{ZJ07} boundary conditions. As a corollary we obtain the normalisation, or sum rule, for the case of two open boundaries. The normalisation is important as the O($n=1$) groundstate can be interpreted as a probability distribution function. In contrast to other boundary conditions, there is as yet no combinatorial interpretation of the normalisation for two open boundaries.

\section{Two-boundary Temperley-Lieb algebra}
Sklyanin's double row transfer matrix \cite{Skly88} of the O($n$) model, and consequently the Hamiltonian, can be expressed in terms of algebraic generators satisfying a Temperley-Lieb algebra, see e.g. \cite{DF05,PearceRZ06}. The particular version of the Temperley-Lieb algebra which is needed depends on the imposed boundary conditions. In this paper we will consider the O($n$) model on a strip with open boundaries on both sides, which can be described in terms of the two-boundary Temperley-Lieb algebra \cite{GierN07}. Models with two reflecting or diagonal boundaries, as well as with mixed boundaries were studied in \cite{DF05,ZJ07,GierP07}.

\begin{defn}
\label{def:2BTL} The two-boundary Temperley-Lieb (2BTL) algebra, or Temperley-Lieb algebra of type $BC_L$, is the algebra over $\mathbb{Z}$ defined in terms of generators $e_i$, $i=0,\ldots,L,\,$ satisfying the relations
\begin{align}
& \begin{array}{@{}l}
e_i^2 =  e_i, \label{ei2a}\\
e_ie_{i\pm1}e_i = e_i, \\
\end{array}\Bigr\} \quad \text{for}\quad i=1,\ldots,L-1,\\
&e_0^2 =  e_0,\qquad e_L^2 = e_L, \non
\end{align}
and commuting otherwise.
\end{defn}
The 2BTL can be generalised to include parameters in the quadratic relations \cite{GierN07}. In particular, the quadratic relations in \eqref{ei2a} can be generalised to
\be
e_i^2=-(q+q^{-1}) e_i,
\ee
for some parameter $q$.\footnote{An even further generalisation is posssible but will not concern us here.} Unless stated otherwise, throughout this paper we will fix $q$ to be the third root of unity,
\be
q=\e^{2\pi\ii/3}.
\label{qspec}
\ee
The Hamiltonian of the dense O($n=1$) loop model with open boundaries can now be expressed as the following operator,
\be
H = c_1(1-e_0) + c_2(1-e_L) + \sum_{j=1}^{L-1} (1-e_j),
\label{eq:ham}
\ee
where $c_1,c_2\in \mathbb{R}$. 

The 2BTL is infinite dimensional, and it was shown in \cite{GierN07} that all finite dimensional irreducible representations satisfy two additional relations, which we will describe now. First we define two (un-normalised) idempotents $I_1$ and $I_2$ as follows:
\begin{itemize}
\item{$L$ even}
\begin{align}
&I_1=e_1e_3\cdots e_{L-1}, && I_2=e_0 e_2 e_4\cdots e_{L-2} e_L.
\label{eqn:IdempotentsEven}
\end{align}
\item{$L$ odd}
\begin{align}
&I_1 = e_1 e_3 \cdots e_{L-2} e_L, && I_2 = e_0 e_2 \cdots e_{L-1}.
\label{eqn:IdempotentsOdd}
\end{align}
\end{itemize}%
The double quotient of the 2BTL algebra has the additional relations:
\begin{align}
&I_1 I_2 I_1 = b I_1, && I_2 I_1 I_2 = b I_2,
\label{eqn:DoubleQuotient}
\end{align}
where $b$ is an additional parameter. 

In the rest of this paper we will restrict to the case $b=1$ so that the 2BTL has a one dimensional representation $\rho$ defined by 
\be
\rho: e_i\mapsto 1,\qquad i=0,1,\ldots, L.
\label{onedim}
\ee
In particular we have that $\rho(H)=0$, hence $0$ is an eigenvalue of $H$ in any faithful representation. In fact, because the eigenvalues of $e_i$ are $0$ and $1$, for $c_1,c_2\geq 0$ the lowest eigenvalue of $H$ is $0$ and corresponds to the groundstate of the O$(n=1)$ loop model. Moreover, by the Perron-Frobenius theorem, the groundstate is unique as $\mathbb{I}-H$ is a stochastic matrix.

The 2BTL algebra has two distinguished representations, both of dimension $2^L$ \cite{GierN07}. One representation is in the tensor product space $(\mathbb{C}^2)^{\otimes L}$, giving rise to the quantum XXZ spin chain with non-diagonal boundary conditions on both sides. In this representation the Temperley-Lieb generators can be expressed in terms of the Pauli matrices $\sigma^{\rm x}$, $\sigma^{\rm y}$ and $\sigma^{\rm z}$, and take the form
\begin{align}
e_i &\mapsto \frac12\left( \sigma_i^{\rm x}\sigma_{i+1}^{\rm x} + \sigma_i^{\rm y}\sigma_{i+1}^{\rm y} + \cos\gamma \left( \sigma_i^{\rm z}\sigma_{i+1}^{\rm z}-1\right) + \ii\sin\gamma  \left( \sigma_i^{\rm z}- \sigma_{i+1}^{\rm z}\right)\right),\non\\
e_0 &\mapsto \frac{1}{2\sin\gamma}\left( -\sin\beta\ \sigma_1^{\rm x} + \cos\beta\ \sigma_1^{\rm y} +\ii \cos 2\gamma\ \sigma_1^{\rm z} - \sin2\gamma\right),\\
e_L &\mapsto \frac{1}{2\sin\gamma}\left( \sin(\beta+\gamma)\ \sigma_L^{\rm x} - \cos(\beta+\gamma)\ \sigma_L^{\rm y} - \ii \cos 2\gamma\  \sigma_1^{\rm z} - \sin2\gamma\right),\non
\end{align}
where we have used the notation $q=\e^{\ii\gamma}$. If we specialise $b=-(q+q^{-1})$, this representation is valid for generic values of $q$. The expressions above furthermore contain an additional angle $\beta$, which, due to the rotational symmetry in the spin $x-y$ plane, is a free gauge parameter on which the spectrum of $H$ does not depend. The choice \eqref{qspec} of $q$ as the third root of unity corresponds to anistropy $\Delta:=\cos\gamma=-1/2$ in the XXZ chain. 

Another representation of the 2BTL algebra is in a space of link patterns, which we will describe in the next section. This representation is relevant for the O($n$) model with open boundaries \cite{MitraNGB04,JacobsS07}. The particular choice \eqref{qspec} of $q$ in this setting corresponds to $n=1$.

\subsection{Action on link patterns}
\begin{figure}[h]
\centerline{
\begin{picture}(200,200)
\put(0,0){\includegraphics[width=200pt]{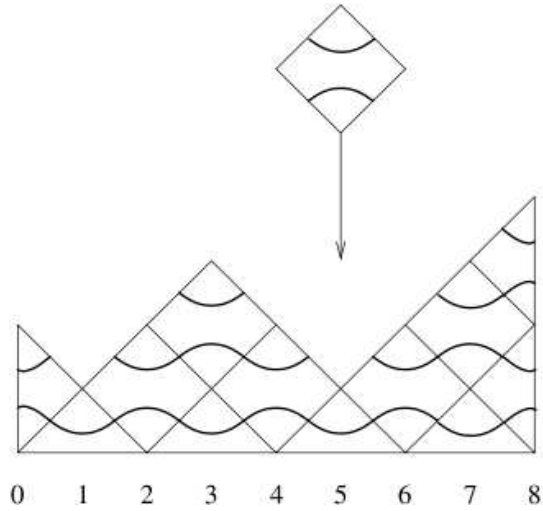}}
\end{picture}}
\caption{Graphical depiction of action of $e_5$ on the anchored cross-path $(2,1,2,3,2,{\ir 1},2,3,4)$ or, equivalently, the link pattern $)((){\ir )(}(($. The result will be the path $(2,1,2,3,2,{\ir 3},2,3,4)$, or link pattern $)((){\ir ()}(($.}
\label{e5onpath}
\end{figure}

The 2BTL algebra has a graphical loop representation in a space spanned by link patterns (sometimes called connectivities) or, equivalently, the space spanned by (a variant of) anchored cross paths \cite{Pyatov04}. 
\begin{defn}
A \textbf{link pattern} is a non-crossing matching of the integers $0,1,\ldots,L+1$. The matching between the integers $1,\ldots, L$ is pairwise, whereas $0$ and $L+1$ may be matched with, or connected to, an arbitrary number of other integers. The integers $0$ and $L+1$ are respectively referred to as the left and right boundary.
\end{defn}
\begin{defn}
An \textbf{anchored cross-path} is a sequence of integers $(h_0,h_1,\ldots,h_L)$ such that $h_{i+1}=h_i \pm 1$ and $\min(h_i) \in\{0,1\}$.
\end{defn}
Before describing these representations more precisely, we present an intuitive picture in Figure~\ref{e5onpath} using the well known graphical depiction of $e_i$ as a tilted square decorated with small loop segments. Multiplication in the 2BTL algebra corresponds to vertical concatenation of pictures. 

We now give a more precise description of the link pattern representation.
Link patterns can be conveniently described by a sequence of opening `(' and closing `)' parentheses.
If site $i$ is matched with, or connected to, site $j$ we put an opening parenthesis `(' at $i$ and a closing parenthesis `)' at $j$. If site $i$ is connected to the left boundary we place a closing parenthesis at $i$. Likewise, if a site $i$ is connected to the right boundary we place an opening parenthesis at $i$. The link patterns for $L=2$ are thus given by
\be
)),\quad )(,\quad (),\quad ((.
\ee
which respectively mean that (i) the two sites are connected to the left boundary, (ii) the first is connected to the left boundary while the second is connected to the right boundary, (iii) the two sites are connected to each other, and (iv) both are connected to the right boundary. Because we can independently place an opening or closing parenthesis at each site, the dimension of the space ${\rm LP}_L$ of link patterns of size $L$ for the two-boundary Temperley-Lieb algebra is 
\be
\dim {\rm LP}_L = 2^L.
\ee

The generator $e_i$ $(i=1,\ldots,L-1)$ acts between positions $i$ and $i+1$ on a link pattern $\alpha\in {\rm LP}_L$ in the following way: If site $i$ is connected to $k$ and site $i+1$ to $l$, $e_i$ connects $i$ with $i+1$ and $k$ with $l$. Here, $k, l\in\{0,\ldots,L+1\}$, but if they both correspond to a boundary ($0$ or $L+1$), then the connection from $k$ to $l$ is disregarded in the final picture. We now describe the action of the generator $e_0$. It connects $1$ to the left boundary, and if $1$ was previously connected to $i\neq 0,L+1$, then $e_0$ also connects $i$ to the left boundary. If $1$ was connected to the left boundary, then $e_0$ acts as the identity. The action of the generator $e_L$ is similar to $e_0$. It connects $L$ to the right boundary, and if $L$ was previously connected to $i\neq 0,L+1$, then $e_L$ also connects $i$ to the right boundary. If $L$ was connected to the right boundary, then $e_L$ acts as the identity.

The representation on anchored cross paths is completely equivalent to that on link patterns. The loop decorations on the tiles define a link pattern in the obvious way, as in Figure~\ref{e5onpath}. The action of a generator follows then from sticking the corresponding tile to a path, connecting all the loop decorations resulting in a new link pattern, and then replacing the picture with the minimal path with the same such link pattern. 

As stated below \eqref{onedim}, the Hamiltonian (\ref{eq:ham}) has a positive spectrum and a unique ground-state energy $E_0=0$ in ${\rm LP}_L$. We will be interested in the corresponding right eigenvector $\ket{\Psi}$ as a function of the parameters $c_1$ and $c_2$,
\be
H\ket{\Psi(c_1,c_2)} =0. 
\label{Hpsi}
\ee
As shown by Di Francesco and Zinn-Justin for other types of boundary conditions, it is possible to derive exact closed form expressions for certain properties of $\ket{\Psi}$ for \textit{finite} system sizes. For example, we will show that the normalisation $Z=\bra{\Psi} \Psi\rangle$, which in the link pattern representation is equal to the sum over all components of $\ket{\Psi}$, can be expressed as a product of four determinants. To achieve this one needs to generalise the eigenvalue problem \eqref{Hpsi} by considering an inhomogeneous model. This is done via the transfer matrix formalism which will be described in the next section.

\section{Transfer matrix}
In order to to define the transfer matrix we will first introduce the operators $\check{R}$ and $\check{K}$, as well as their unchecked versions. We furthermore list some useful properties wich we will need in later sections. Throughout the following we will use the notation $[z]$ for 
\be
[z] = z-z^{-1}.
\ee

\subsection{Baxterisation}

The Baxterised elements $\check R_i(z)$, and the boundary Baxterised elements $\check K_0(z,\zeta)$ and $\check K_L(z,\zeta)$ of the Temperley-Lieb algebras are defined as
\be
\lb{R_z}
\begin{split}
\check R_i(z)&=\,\frac{[q/z]-[z]\,e_i}{[qz]}\,,\\
\check K_i(z,\zeta)& =\,\frac{k(z,\zeta)-[q][z^2]e_i}{k(1/z,\zeta)},\quad i=0,L,\\
\end{split}
\ee
where $k(z,\zeta)$ is given by
\be
k(z,\zeta)=[z/q\zeta][z\zeta/q].
\label{kdef}
\ee
The parameter $z$ is called the spectral parameter. In addition, each boundary element can be equipped with an additional free parameter $\zeta$. We thus will have two such parameters available, and they will be related to the coefficients $c_1$ and $c_2$ in \eqref{eq:ham} by
\be
c_i = \frac{(q-1/q)^2}{(q^2-\zeta_i^2)(1-q^{-2}\zeta_i^{-2})} =  \frac{3}{1+\zeta_i^2+\zeta_i^{-2}}\ .
\ee
The Baxterised elements obey the usual Yang-Baxter and reflection equations with spectral parameters:
\begin{align}
\lb{eq:RKcheckYbeReflect}\check R_i(z)\check R_{i+1}(zw)\check R_i(w) &= \check R_{i+1}(w)\check R_i(zw)\check R_{i+1}(z),\nonumber\\
\check K_0(z,\zeta)\check R_{1}(zw)\check K_0(w,\zeta)\check R_1(w/z) &= \check R_{1}(w/z)\check K_0(w,\zeta)\check R_{1}(zw)\check K_0(z,\zeta),\\
\check K_L(z,\zeta)\check R_{L-1}(zw)\check K_L(w,\zeta)\check R_{L-1}(w/z) &= \check R_{L-1}(w/z)\check K_L(w,\zeta)\check R_{L-1}(zw)\check K_L(z,\zeta).\non
\end{align}
They furthermore satisfy the unitarity relations
\be
\lb{eq:RKcheckId}
\begin{split}
\check R_i(z)\check R_i(1/z)&=1,\\
\check K_i(z,\zeta)\check K_i(1/z,\zeta)&=1,\quad i=0,L.
\end{split}
\ee

The relations above can be easily checked using the algebraic rules \eqref{ei2a}, or using a graphical notation like the one in Figure~\ref{e5onpath}. We now introduce a graphical definition of the Baxterised elements, using the planar Temperley-Lieb-Jones algebra \cite{Jones99}, which we will be able to use in a more general context than Figure~\ref{e5onpath}. We thus define the R-operator $R(z,w)$ to be the following linear combination of pictures:
\begin{align}
R(z,w)&=\frac{[qw/z]}{[qz/w]}\quad\raisebox{-13pt}{\includegraphics[height=30pt]{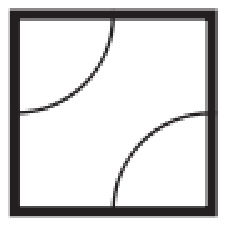}}-\;\frac{[z/w]}{[qz/w]}\quad\raisebox{-13pt}{\includegraphics[height=30pt]{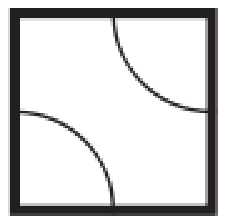}},
\end{align}
and will graphically abbreviate $R(z,w)$ by
\be
R(z,w) =\raisebox{-33pt}{\includegraphics[height=70pt]{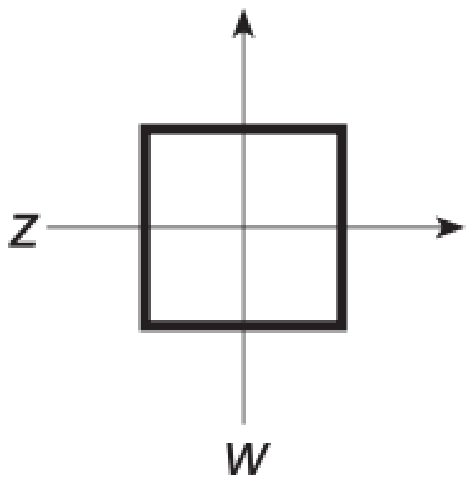}}.
\ee
Note that we can use this picture in any orientation, as the arrows uniquely determine how the spectral parameters $z$ and $w$ enter in $R$. Likewise, we define the boundary K-operators by
\begin{align}
K_0(w,\zeta)&=\frac{k(qw,\zeta)}{k(1/qw,\zeta)}\quad\raisebox{-18pt}{\includegraphics[height=40pt]{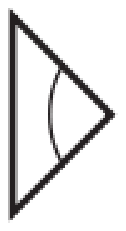}}\;-\frac{[q][q^2w^2]}{k(1/qw,\zeta)}\quad\raisebox{-18pt}{\includegraphics[height=40pt]{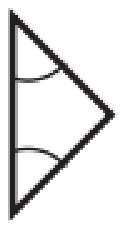}}\non\\
&=\raisebox{-23pt}{\includegraphics[height=50pt]{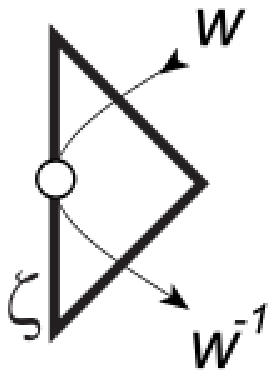}},\\
K_L(w,\zeta)&=\frac{k(w,\zeta)}{k(1/w,\zeta)}\quad\raisebox{-18pt}{\includegraphics[height=40pt]{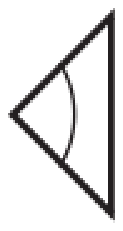}}\;-\frac{[q][w^2]}{k(1/w,\zeta)}\quad\raisebox{-18pt}{\includegraphics[height=40pt]{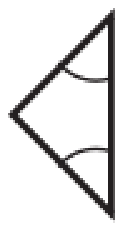}}\non\\
&=\raisebox{-23pt}{\includegraphics[height=50pt]{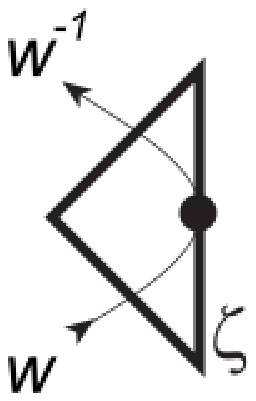}}.
\end{align}
For simplicity we will sometimes only draw the lines carrying the arrows and spectral parameters. The Baxterised elements $R$, $K_0$ and $K_L$ will be used to define the transfer matrix of the system. 

The unitarity relations \eqref{eq:RKcheckId} for $R$  and $K$ can be graphically depicted as
\be
\raisebox{-35pt}{\includegraphics[height=60pt]{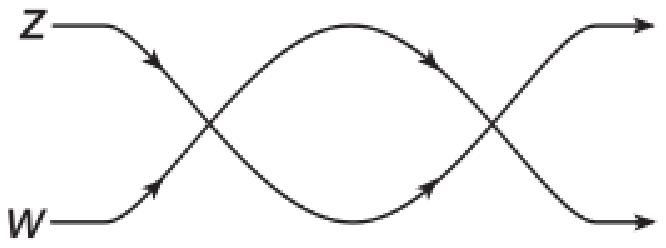}}\quad=\quad\raisebox{-35pt}{\includegraphics[height=60pt]{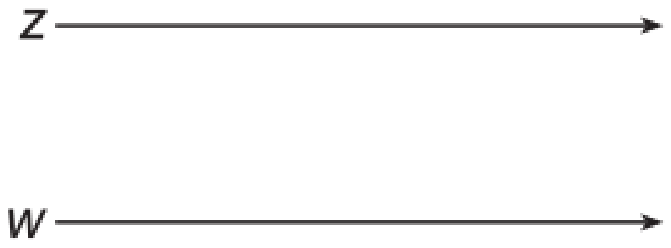}}\;,
\ee
and
\be
\raisebox{-45pt}{\includegraphics[height=100pt]{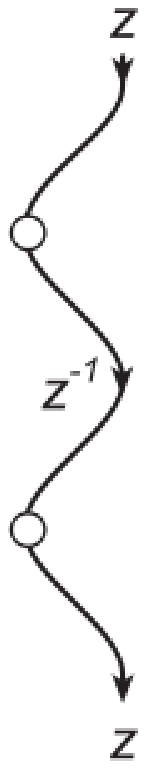}}\quad=\quad\raisebox{-45pt}{\includegraphics[height=100pt]{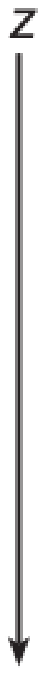}}\;,\qquad\qquad\qquad\raisebox{-45pt}{\includegraphics[height=100pt]{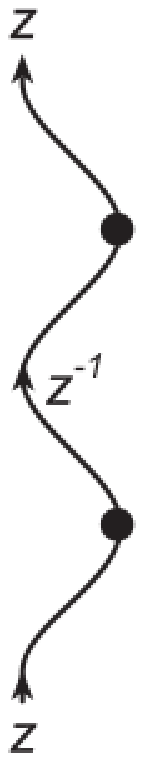}}\quad=\quad\raisebox{-45pt}{\includegraphics[height=100pt]{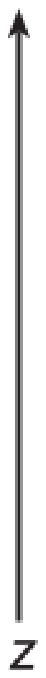}}\;.
\ee
In addition, the Yang-Baxter and reflection equations \eqref{eq:RKcheckYbeReflect} can be written as
\be
\raisebox{-45pt}{\includegraphics[height=80pt]{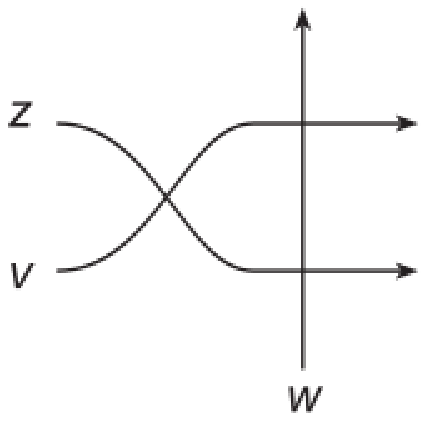}}\quad=\quad\raisebox{-45pt}{\includegraphics[height=80pt]{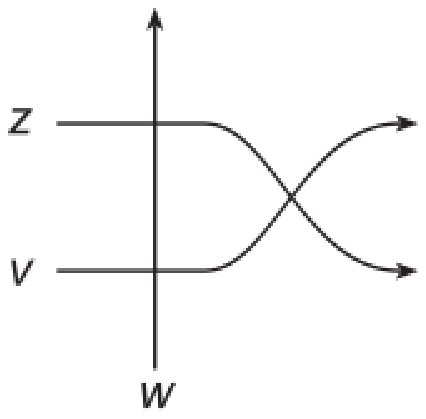}}\;,
\ee
and
\be
\raisebox{-45pt}{\includegraphics[height=100pt]{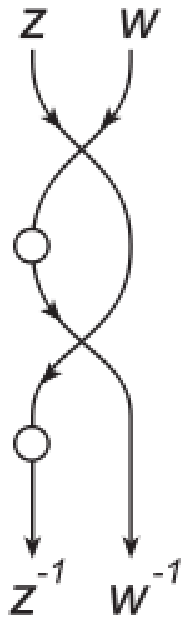}}\quad=\quad\raisebox{-45pt}{\includegraphics[height=100pt]{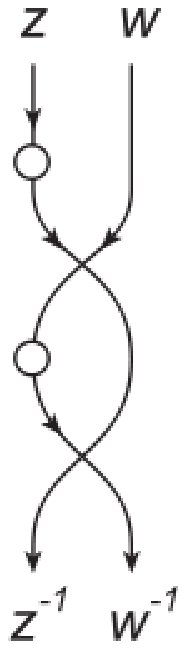}},\qquad\qquad\qquad\raisebox{-45pt}{\includegraphics[height=100pt]{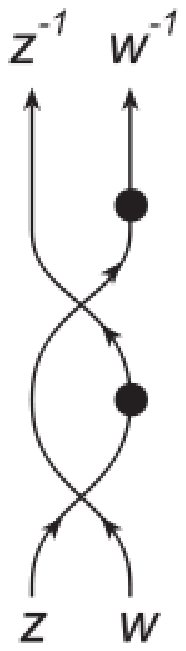}}\quad=\quad\raisebox{-45pt}{\includegraphics[height=100pt]{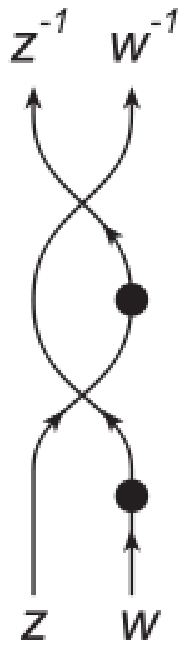}}\;.
\ee
We furthermore note the crossing relation satisfied by $R$, i.e. $R(z,w)=R(qw,z)$ (recall that $q=\e^{2\pi\ii/3}$), which graphically reads
\be
\raisebox{-35pt}{\includegraphics[height=60pt]{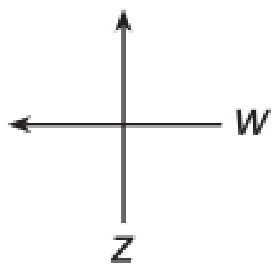}}\quad=\quad\raisebox{-35pt}{\includegraphics[height=60pt]{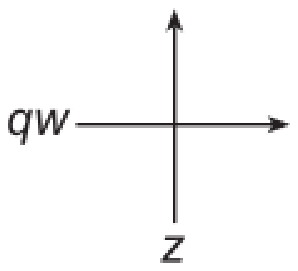}}\;.
\label{crossing}
\ee
The crossing relation \eqref{crossing} can be made to hold for generic values of $q$ if $R$ is normalised appropriately \cite{JimboM94}.

\subsection{Transfer Matrix}
\lb{sec:transfermatrix}

Using the definitions in the previous section we define Sklyanin's double row transfer matrix $T_L(w)=T_L(w;z_1,\ldots,z_L;\zeta_1,\zeta_2)$ pictorially in the following way, see \cite{Skly88,DF05,PearceRZ06},
\be
T_L(w) = \raisebox{-60pt}{\includegraphics[height=120pt]{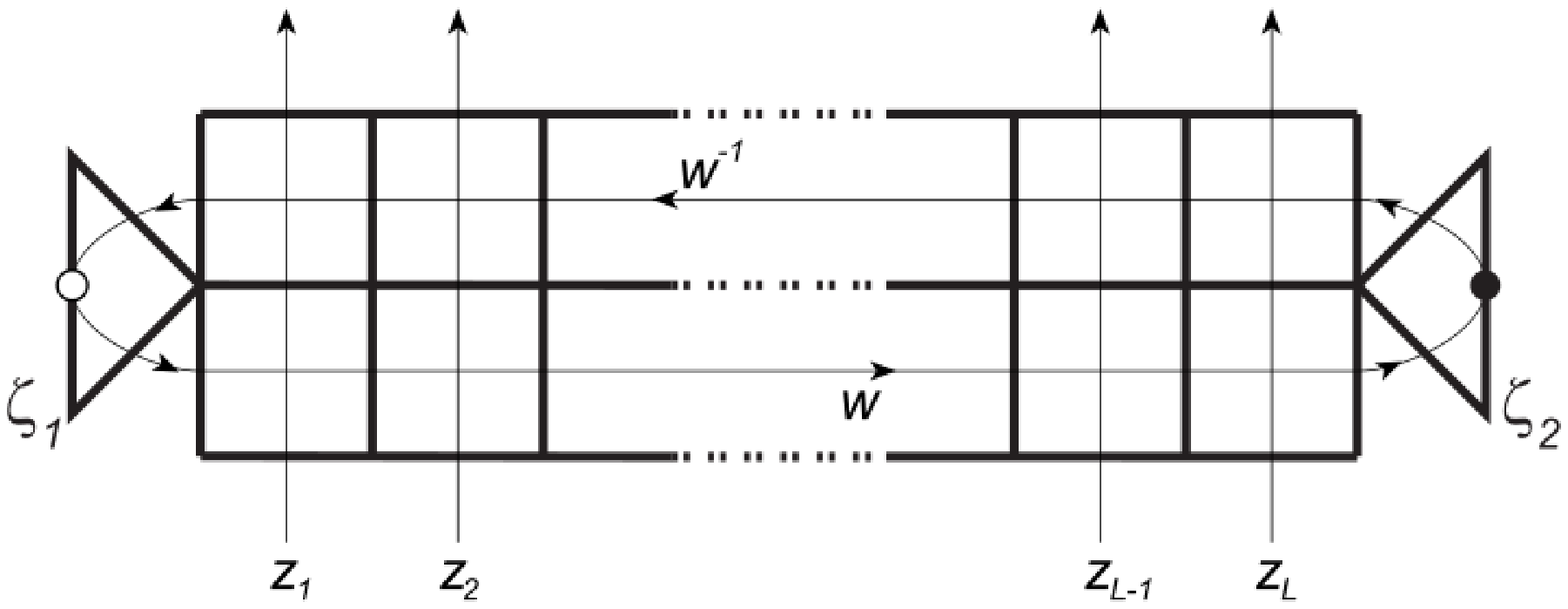}}\;.
\ee
As is well known, the Yang-Baxter and reflection equations \eqref{eq:RKcheckYbeReflect} imply that $T(w)$ forms a commuting family of transfer matrices, i.e.
\be
[T_L(v),T_L(w)]=0,
\ee
and hence define an integrable lattice model. Following \cite{DFZJ04,DF05}, we note that the Yang-Baxter and reflection equations \eqref{eq:RKcheckYbeReflect} also immediately imply the following interlacing conditions of the transfer matrix with $\check R_i$,  $\check K_0$ and  $\check K_L$: 
\begin{align}
\check R_i(z_i/z_{i+1})T_L(w;z_1,\ldots,z_L) &= T_L(w;z_1,\ldots,z_{i+1},z_i,\ldots,z_L)\check R_i(z_i/z_{i+1}),\non\\
\lb{eq:interlace}\check K_0(z_1^{-1},\zeta_1)T_L(w;z_1,\ldots,z_L) &= T_L(w;z_1^{-1},z_2,\ldots,z_L)\check K_0(z_1^{-1},\zeta_1),\\
\check K_L(z_L,\zeta_2)T_L(w;z_1,\ldots,z_L) &= T_L(w;z_1,\ldots,z_{L-1},z_L^{-1})\check K_L(z_L,\zeta_2).\non
\end{align}

Due to the existence of the one dimensional representation \eqref{onedim}, the transfer matrix has an eigenvalue equal to $1$, and a corresponding eigenvector $\ket\Psi_L$ defined by
\be
\lb{eq:evec1}T_L(w;z_1,\ldots,z_L)\ket{\Psi(z_1,\ldots,z_L)}_L=\ket{\Psi(z_1,\ldots,z_L)}_L,
\ee
where $\ket\Psi_L$ depends on $z_i$ but not on $w$. In the homogeneous limit $z_i\rightarrow 1$, the transfer matrix $T_L$ becomes the probability transition matrix of the stochastic raise and peel model \cite{PearceRGN02,Pyatov04}, for which the steady state eigenvector \eqref{eq:evec1} is unique by the Perron-Frobenius theorem. We will assume that the eigenvector remains unique for generic values for $z_1,\ldots,z_L$. The vector $\ket\Psi_L$ is the groundstate eigenvector for the O($n=1$) loop model with open boundaries. In Section~\ref{se:qKZ} we will use the interlacing conditions \eqref{eq:interlace} to rewrite \eqref{eq:evec1} in a form which is known as the $q$-deformed Knizhnik-Zamolodchikov equation. This will allow us to obtain an explicit characterisation of $\ket\Psi_L$ for finite $L$. We will in particular be able to derive a closed form expression for the normalisation $Z_L=\bra\Psi \Psi\rangle_L$. In order to do so we need a recursion relation for $\ket\Psi_L$, which we will discuss first.

\subsection{Recursion}

Let $\varphi_i$ denote the map that sends site $j$ to $j+2$ for $j\geq i$ in a link pattern, and then inserts a link from site $i$ to $i+1$, thus creating a link pattern of size two greater. For example,
\be
\varphi_3:\quad )(()((\quad \mapsto\quad )({\ir ()}()((.
\ee
In Appendix~\ref{ap:recursion} we prove that the transfer matrix satisfies the following identity:
\be
\lb{eq:Tphi}T_L(w;z_1,\ldots,z_{i+1}=qz_i,\ldots,z_L)\circ\varphi_i=\frac{[q/z_iw][q^2z_i/w]}{[q^2z_iw][qw/z_i]}\ \varphi_i\circ T_{L-2}(w;\ldots,z_{i-1},z_{i+2},\ldots).
\ee
A similar relation was proved in \cite{DF05} and for the case of periodic boundary conditions in \cite{DFZJ04}. Property \eqref{eq:Tphi} will be used later, in particular in conjunction with $q=\e^{2\pi\ii/3}$ when the proportionality factor equals 1.

Likewise one can prove that at the boundaries, and for $q=\e^{2\pi\ii/3}$, the transfer matrix satisfies
\be
T_L(w;z_1=q\zeta_1,\ldots,z_L;\zeta_1,\zeta_2) \circ \varphi_0 = \varphi_0 \circ T_{L-1}(w;z_2,\ldots,z_L;q\zeta_1,\zeta_2),
\label{eq:0boundrecur}
\ee
where $\varphi_0$ is the map that sends site $j$ to site $j+1$, and inserts a $)$ on the first site; and
\be
T_L(w;z_1,\ldots,z_L=\zeta_2/q;\zeta_1,\zeta_2) \circ \varphi_L = \varphi_L \circ T_{L-1}(w;z_1,\ldots,z_{L-1};\zeta_1,\zeta_2/q),
\label{eq:Lboundrecur}
\ee
where $\varphi_L$ is the map that inserts a $($ on the last site.

\section{The $q$-Knizhnik-Zamolodchikov equation}
\label{se:qKZ}
The groundstate eigenvalue equation \eqref{eq:evec1} for the inhomogeneous transfer matrix of the O(${n=1}$) model is equivalent to a particular instance of the $q$KZ equation with $q=\e^{2\pi\ii/3}$. This connection will provide a handle for an explicit analysis of the groundstate eigenvector of the O($n=1$) Hamiltonian for finite system size $L$. We will first describe the $q$KZ equation for open boundaries (corresponding to type BC in Dynkin diagram classification), and then prove the equivalence with the transfer matrix eigenvalue equation \eqref{eq:evec1}.

We consider a linear combination $\ket\Xi$ of states $\ket \alpha$ labeled by link patterns:
\be
\ket{\Xi(z_1,\ldots,z_{L})} = \sum_{\alpha}
\xi_\alpha(z_1,\ldots,z_{L}) \ket{\alpha}.
\label{Xi}
\ee
Here $\ket \alpha$ runs over the set of anchored cross paths (or link patterns) of size $L$, and the coefficient functions $\xi_\alpha$ are polynomials in $L$ variables with coefficients which are functions of $q$ and a new parameter $s$, which we regard as complex parameters,
\be
\xi_\alpha \in \mathbb{C}[z_1^2,\ldots,z_L^2].
\ee
The $q$-Knizhnik-Zamolodchikov equation \cite{smirnov,FR,Cher} is a system of finite difference equations on the vector $\ket\Xi$.
For open boundary conditions they can be written as \cite{DF05,ZJ07},\footnote{We write the equations in a form used by Smirnov \cite{smirnov}, which imply the more commonly used form of Frenkel and Reshetikhin \cite{FR}.}
\begin{align}
\check R_i(z_i/z_{i+1}) \ket\Xi &= \pi_i \ket\Xi,\qquad \forall\, i\in\{1,\ldots,L-1\},\non
\\
\check K_0(1/z_1,\zeta_1) \ket\Xi  &= \pi_0 \ket\Xi,
\label{qKZTL_TypeC2}\\
\check K_L(s z_L,s\zeta_2) \ket\Xi &= \pi_L\ket\Xi.\non
\end{align}
The operators $\check R_i(z)$  are the Baxterised elements of the Temperley-Lieb algebra, and $\check K_0$ and $\check K_L$ are the boundary Baxterised elements from \eqref{R_z}. The operators $\check R_i(z_i/z_{i+1})$, $\check K_0(1/z_1,\zeta_1)$ and $\check K_L(sz_L,s\zeta_2)$  act on paths (link patterns) $\ket \alpha$, whereas the operators $\pi_i$ ($i=0,\dots , L$) act on the coefficient functions $\xi_\alpha$ only;
\begin{align}
\pi_i \xi(\ldots,z_i,z_{i+1},\ldots) &= \xi(\ldots,z_{i+1},z_{i},\ldots),\nonumber\\
\pi_0\xi(z_1,\ldots)&= \xi(1/z_1,\ldots),\label{pi0}\\
\pi_L\xi(\ldots,z_L)&= \xi(\ldots,1/s^2z_L).\label{piL}\nonumber
\end{align}
%
For later convenience, we note that the $q$KZ equations can be rewritten in the following way,
\be
\lb{eq:qKZea}
e_i \ket\Xi = -a_i \ket\Xi\qquad (i=0,\ldots,L),
\ee
where
\begin{align}
a_i &= (\pi_i+1) \frac{[z_i/qz_{i+1}]}{[z_i/z_{i+1}]},\non\\
a_0 &= (\pi_0+1) \frac{k(1/z_1,\zeta_1)}{[q][z_1^2]},\\
a_L &= -(\pi_L+1) \frac{k(sz_L,s\zeta_2)}{[q][s^2z_L^2]},\non
\end{align}
where $k$ was defined in \eqref{kdef}. The operators $a_i$ ($i=0,\ldots,L$) satisfy the relations of the affine Hecke algebra of type C \cite{GierP07}, as well as those of the Hecke algebra of type BC.

\subsection{Equivalence with the transfer matrix eigenvalue equation}
The $q$KZ equation has polynomial solutions for special values of $s$ and $q$.\footnote{Interesting recent developments \cite{KT} relate polynomial solutions of the $q$KZ equation associated to $U_q(sl_n)$, to the polynomial representation of the double affine Hecke algebra \cite{Noumi95,Sahi}. These solutions can be expressed in terms of Macdonald polynomials with specialised parameters \cite{FeiginJMM}.} At special values (namely $q=\e^{2\pi\ii/3}, s^4=1$), the groundstate eigenvector $\ket\Psi_L$ of the transfer matrix also satisfies the $q$KZ equation. We prove this by acting on both sides of the eigenvector equation \eqref{eq:evec1} with the elements $\check R(z_i/z_{i+1})$, $\check K_0(1/z_1,\zeta_1)$ and $\check K_L(sz_L,s\zeta_2)$,
\begin{align}
\check R_i(z_i/z_{i+1})\ket\Psi_L&=\check R_i(z_i/z_{i+1})T_L(w;z_1,\ldots,z_L)\ket\Psi_L\non\\
&=T_L(w;z_1,\ldots,z_{i+1},z_i,\ldots,z_L)\check R_i(z_i/z_{i+1})\ket\Psi_L.
\end{align}
Acting on both sides with $\pi_i$, we obtain
\be
T_L(w;z_1,\ldots,z_L)(\pi_i\check R_i(z_i/z_{i+1})\ket{\Psi(z_1,\ldots,z_L)}_L=\pi_i\check R_i(z_i/z_{i+1})\ket{\Psi(z_1,\ldots,z_L)}_L,
\ee
and since the eigenvector in \eqref{eq:evec1} is unique, this implies that
\be
\lb{eq:prop1}\pi_i\check R_i(z_i/z_{i+1})\ket{\Psi(z_1,\ldots,z_L)}_L=\beta_i(z_1,\ldots,z_L) \ket{\Psi(z_1,\ldots,z_L)}_L,
\ee
where $\beta_i$ is some rational function. Multiplying both sides of \eqref{eq:prop1} with $\pi_i\check R_i(z_i/z_{i+1})$, and using the identity in \eqref{eq:RKcheckId}, it follows that
\be
\ket{\Psi(z_1,\ldots,z_L)}_L=\beta_i(\ldots,z_{i+1},z_i,\ldots) \pi_i \check R_i(z_i/z_{i+1})\ket{\Psi(z_1,\ldots,z_L)}_L,
\label{Psibeta}
\ee
so that 
\be
\beta(\ldots,z_i,z_{i+1},\ldots)\beta(\ldots,z_{i+1},z_i,\ldots)=1.
\label{betarecur} 
\ee
Because we may assume that the components of $\ket\Psi_L$ do not have a common factor, equation \eqref{Psibeta} implies that the numerator of $\beta_i$ must be a constant, or that $1/\beta_i$ is a polynomial. But then \eqref{betarecur} implies that $\beta_i$ is in fact a constant such that $\beta_i^2=1$. The sign is fixed to $+1$ by setting $z_{i+1}=z_i$, so we finally obtain 
\be
\check R_i(z_i/z_{i+1})\ket{\Psi(z_1,\ldots,z_L)}_L=\pi_i \ket{\Psi(z_1,\ldots,z_L)}_L.
\ee
Similarly, we find that
\be
\begin{split}
\check K_0(1/z_1,\zeta_1)\ket{\Psi(z_1,\ldots,z_L)}_L &= \pi_0\ket{\Psi(z_1,\ldots,z_L)}_L,\\
\check K_L(sz_L,s\zeta_2)\ket{\Psi(z_1,\ldots,z_L)}_L &= \pi_L\ket{\Psi(z_1,\ldots,z_L)}_L,
\end{split}
\ee
where proof of the last equation makes use of the fact that when $s^4=1$, $\check K_L(sa,sb)=\check K_L(a,b)$ and $R(s^2z,w)=R(z,w)$.

\subsection{Example $L=2$}
\label{se:L=2}
In this section we work out an example of a solution of the $q$KZ equation for $L=2$ which also solves the eigenvalue equation \eqref{eq:evec1}. We work in the link pattern basis $\{ )), )(, (), ((\}$, which is equivalent to $(2,1,0)$, $(2,1,2)$, $(0,1,0)$ and $(0,1,2)$ in the path representation. In this case, there are twelve equations resulting from the $q$KZ equations. Defining 
\be
s_i=q+q^{-1}-a_i,
\label{sidef}
\ee
these equations can be written as
\be
\begin{split}
0&= a_0 \psi_{()} = a_0 \psi_{((},\\
\psi_{()} &= s_0 \psi_{))}, \qquad \psi_{((}= s_0\psi_{)(}\ , 
\end{split}
\label{eq:froma0}
\ee

\be
\begin{split}
0&= a_2 \psi_{()} = a_2 \psi _{))}, \\
\psi_{()}&= s_2 \psi_{((}, \qquad \psi_{))}= s_2\psi_{)(}\ ,
\end{split}
\label{eq:froma2}
\ee

\be
\begin{split}
0&= a_1 \psi_{))} = a_1 \psi _{)(} = a_1 \psi_{((}\ , \\
& s_1\psi_{()} = \psi_{))} + \psi_{((} + \psi_{)(}\ .
\end{split}
\label{eq:froma1}
\ee
It is an easy consequence of the equation $a_0\psi_{()}=0$ that if $\psi_{()}\neq 0$, it should contain a factor $k(z_1,\zeta_1)$. Similar conditions hold for the vanishing of $\psi$  when acted upon by $a_1$ and $a_2$:\bigskip
\begin{itemize}
\item[i.] $\psi_{()}$ and $\psi_{((}$ vanish or contain a factor $k(z_1,\zeta_1)$, the remainder being invariant under $z_1 \leftrightarrow 1/z_1$.
\item[ii.] $\psi_{()}$ and $\psi_{))}$ vanish or contain a factor $k(1/sz_2,s\zeta_2)$, the remainder being invariant under $sz_2\leftrightarrow 1/sz_2$.
\item[iii.] $\psi_{))}$, $\psi_{)(}$ and $\psi_{((}$ vanish or contain a factor $[qz_1/z_2]$, the remainder being a symmetric function in $z_1$ and $z_2$.
\end{itemize}

\subsubsection{Solution}
With the known factors from items i., ii. and iii. above, we thus look for a solution of the form
\be
\begin{split}
\psi_{))} &= \prod_{i=1}^2 k(1/sz_i,s\zeta_2)\times [qz_1/z_2][qs^2z_1z_2]\times S(sz_1,sz_2), \\
\psi_{((} &= \prod_{i=1}^2 k(z_i,\zeta_1)\times [qz_1/z_2][q/z_1z_2]\times \widetilde{S}(z_1,z_2).
\end{split}
\ee
where $S(z_1,z_2)$ and $\widetilde{S}(z_1,z_2)$ are symmetric functions invariant under $z_i\leftrightarrow 1/z_i$. The other two components may be determined from
\be
\lb{eq:L2psi2}
\psi_{()} = s_0 \psi_{))},\qquad \psi_{)(} = s_1\psi_{()} - \psi_{))} - \psi_{((}.
\ee
When $s^4=1$, $q=\e^{2\pi\ii/3}$ we find that the solution to equations \eqref{eq:froma0}--\eqref{eq:froma1} can be given explicitly by
\be
\lb{eq:L2S}
\begin{split}
S(z_1,z_2) &= A_2\, \chi_{(1,0,0)}(s^2\zeta_1^2,z_1^2,z_2^2),\\
\widetilde{S}(z_1,z_2) &= A_2\, \chi_{(1,0,0)}(z_1^2,z_2^2,\zeta_2^2),
\end{split}
\ee
where, with an eye on generalisation to arbitrary $L$, we have used the symplectic character $\chi_\lambda$ of degree $\lambda$ defined by
\be
\chi_\lambda(z_1,\ldots,z_L) = \frac{\left|z_i^{\lambda_j+L-j+1}-z_i^{-\lambda_j-L+j-1}\right|_{1\leq i,j\leq L}}{\left|z_i^{L-j+1}-z_i^{-L+j-1}\right|_{1\leq i,j\leq L}},
\label{sympchar}
\ee
and $A_2$ is an arbitrary overall constant. 

\section{Solution for general system size}

As in items i., ii. and iii. of Section~\ref{se:L=2}, for general $L$ we may derive factors for certain components. For each $i$, every link pattern in the LHS of the $q$KZ equation \eqref{eq:qKZea} will have a small link from $i$ to $i+1$ once $e_i$ has acted. The $q$KZ equation then says that $a_i\psi_\alpha=0$ iff $\alpha$ does not have a small link from $i$ to $i+1$. This leads to the following conditions on $\psi_\alpha$:
\begin{itemize}
\lb{psiconditions}
\item[i.] If $\alpha$ does not have a small link from the left boundary to $1$, $\psi_\alpha$ vanishes or contains a factor $k(z_1,\zeta_1)$, the remainder being invariant under $z_1 \leftrightarrow 1/z_1$.
\item[ii.] If $\alpha$ does not have a small link from $L$ to the right boundary, $\psi_\alpha$ vanishes or contains a factor $k(1/sz_L,s\zeta_2)$, the remainder being invariant under $sz_L\leftrightarrow 1/sz_L$.
\item[iii.] If $\alpha$ does not have a small link from $i$ to $i+1$, $\psi_\alpha$ vanishes or contains a factor $[qz_i/z_{i+1}]$, the remainder being a symmetric function in $z_i$ and $z_{i+1}$.
\end{itemize}
Using the above conditions, for general $L$ the component $\psi_{(\cdots(}$ is given by
\be
\psi_{(\cdots(} = \prod_{i=1}^L k(z_i,\zeta_1)\ \prod_{1\leq i<j \leq L} [qz_i/z_j] [q/z_iz_j]\ f_L(z_1,\ldots,z_L),
\ee
where $f_L$ is symmetric and invariant under $z_i\rightarrow 1/z_i$. The majority of the factors in this expression are imposed by the symmetry conditions. Note that with $k$ as defined in \eqref{kdef} we could write this as
\be
\label{eq:psidef1}\psi_{(\cdots(} = \prod_{0\leq i<j \leq L} k(z_j,z_i)\ f_L(z_1,\ldots,z_L),
\ee
by using the notation $z_0=\zeta_1$.

Likewise, we have
\begin{align}
\psi_{)\cdots)} &= \prod_{i=1}^L k(1/sz_i,s\zeta_2)\ \prod_{1\leq i<j \leq L} [qz_i/z_j] [qs^2z_iz_j]\ \tilde f_L(sz_1,\ldots,sz_L),\non\\
\label{eq:psidef2}&= \prod_{1\leq i<j \leq L+1} k(1/sz_i,sz_j)\ \tilde f_L(sz_1,\ldots,sz_L),
\end{align}
if we identify $z_{L+1}=\zeta_2$. Here, $\tilde f_L$ is symmetric and invariant under $sz_i\rightarrow 1/sz_i$. Other components may be derived from the extremal components by acting with products of Baxterised versions of the operators $s_i$, as described in \cite{GierP07}. However, in the case under consideration it is not possible to derive every component of $\ket\Psi$ in this way. In Appendix~\ref{ap:L=3} we explain the reasons for this in detail for the case $L=3$.

By using recursion and degree properties of the general solution, we can find an expression for $\tilde f_L$ and $f_L$. We emphasise again that throughout this section we will use $s^4=1$ and $q=\e^{2\pi\ii/3}$.

\subsection{Recursion}
\subsubsection{Recursion of the eigenstate}

In order to find a recursive definition for all components of $\ket\Psi_L$, we must refer to the recursive property of the transfer matrix described in \eqref{eq:Tphi}. For the remainder of this section, we will suppress the arguments $z_1,\ldots,z_L$ of $T$ and $\ket\Psi_L$ except where detail is needed. The notation $\hat{z}_j$ will mean that $z_j$ is missing from the list $z_1,\ldots,z_L$. When we specify $q$ to be a third root of unity, the factors in \eqref{eq:Tphi} cancel, so we have
\be
T_L(w;z_{i+1}=qz_i)\circ\varphi_i=\varphi_i\circ T_{L-2}(w;\hat z_i,\hat z_{i+1}).
\ee
Now, acting with both sides on the vector $\ket{\Psi(\hat z_i,\hat z_{i+1})}_{L-2}$, we get
\be
T_L(w;z_{i+1}=qz_i)\ \Bigl(\varphi_i \ket{\Psi(\hat z_i,\hat z_{i+1})}_{L-2}\Bigr) = \varphi_i \ket{\Psi(\hat z_i,\hat z_{i+1})}_{L-2}\ ,
\ee
which, by uniqueness of the eigenvector $\ket\Psi_L$, implies
\be
\lb{eq:psiprop}\ket{\Psi(z_{i+1}=qz_i)}_L\propto\varphi_i \ket{\Psi(\hat z_i,\hat z_{i+1})}_{L-2}.
\ee
This is consistent with the properties listed in Section~\ref{se:L=2}, which imply that any coefficient of a link pattern without a small link connecting $i$ and $i+1$ vanishes when $z_{i+1}=qz_i$. Relation \eqref{eq:psiprop} was already proved for subcases of the most general open boundary conditions in \cite{DF05,ZJ07}, and for periodic boundary conditions in \cite{DFZJ04}. It can be shown that the proportionality factor in \eqref{eq:psiprop} takes the same form for each $i$, see Appendix \ref{ap:propsym}. We denote this factor by $p(z_i;z_1,\ldots,\hat z_i,\hat z_{i+1},\ldots,z_L)$.

Likewise, from the boundary recursions \eqref{eq:0boundrecur} and \eqref{eq:Lboundrecur} of the transfer matrix we find that
\be
\begin{split}
\label{eq:boundrecurpsi}
\ket{\Psi(z_1=q\zeta_1;\zeta_1)}_L &=r_0(z_2,\ldots,z_L;\zeta_1) \varphi_0 \ket{\Psi(\hat{z}_1;q\zeta_1)}_{L-1},\\
\ket{\Psi(z_L=\zeta_2/q;\zeta_2)}_L &=r_L(z_1,\ldots,z_{L-1};\zeta_2) \varphi_L \ket{\Psi(\hat{z}_L;\zeta_2/q)}_{L-1},
\end{split}
\ee
where $r_0$ and $r_L$ are proportionality factors analogous to $p$.

\subsubsection{Small size examples}
\label{se:smallsize}
In Section \ref{se:L=2} we presented a minimal degree solution for $L=2$:
\be
\begin{split}
\psi_{((}&=A_2\prod_{0\leq i<j\leq 2}k(z_j,z_i)\ \chi_{(1,0,0)}(z_1^2,z_2^2,\zeta_2^2)\\
\psi_{))}&=A_2\prod_{1\leq i<j\leq 3}k(1/sz_i,sz_j)\ \chi_{(1,0,0)}(s^2\zeta_1^2,s^2z_1^2,s^2z_2^2).
\end{split}
\ee
When we set $z_2=qz_1$, all components except for $\psi_{()}$ vanish. At this point, we have from \eqref{eq:L2psi2}
\begin{align}
\psi_{()}|_{z_2=qz_1}&=(-1-a_0)\psi_{))}|_{z_2=qz_1}\non\\
&=-A_2\ \pi_0\frac{k(1/z_1,\zeta_1)}{[q][z_1^2]}k(1/sz_1,s\zeta_2)k(1/sz_2,s\zeta_2)k(1/sz_1,sz_2)\non\\
&\quad\times\chi_{(1,0,0)}(s^2z_1^2,s^2z_2^2,s^2\zeta_1^2)|_{z_2=qz_1}\non\\
&=-A_2\, k(z_1,\zeta_1)^2k(z_1,\zeta_2)^2,
\label{eq:L2recur}
\end{align}
where we have used the properties $k(s^{2}a,b)=s^{2}k(sa,sb)=k(a,1/b)=k(1/qa,b)=k(a,b)$, the definition of $\chi$ given in \eqref{sympchar}, and the fact that $\psi_{))}$ vanishes when $z_2=qz_1$. Since the solution for $L=0$ is simply a constant scalar (denoted $A_0$), we can easily see that the proportionality factor in \eqref{eq:psiprop} for $L=2$ is equal to
\be
p(z_1)=-\frac{A_2}{A_0}k(z_1,\zeta_1)^2k(z_1,\zeta_2)^2.
\ee

For $L=1$, a minimal degree solution is given by
\be
\begin{split}
\psi_(&=A_1\,k(z_1,\zeta_1)\\
\psi_)&=A_1\,s^2k(1/sz_1,s\zeta_2).
\end{split}
\ee
It is computationally very intensive to compute explicitly the full solution for $L=3$. However, if we restrict to the subset of equations so that $\psi_{)((}$ and $\psi_{))(}$ are not individually determined, but only their sum is, we find
\be
\begin{split}
\psi_{(((}&=A_3\prod_{0\leq i<j\leq 3}k(z_j,z_i)\ \chi_{(1,1,0,0)}(z_1^2,z_2^2,z_3^2,\zeta_2^2) g_3(z_1^2,z_2^2,z_3^2)\\
\psi_{)))}&=A_3\ s^2 \prod_{1\leq i<j\leq 4}k(1/sz_i,sz_j)\ \chi_{(1,1,0,0)}(s^2z_1^2,s^2z_2^2,s^2z_3^2,s^2\zeta_1^2) g_3(s^2z_1^2,s^2z_2^2,s^2z_3^2),
\end{split}
\ee
with $\chi$ as before, and where $g_3$ is symmetric and invariant under $z_i\leftrightarrow 1/z_i$. Imposing the boundary recursions \eqref{eq:0boundrecur} and \eqref{eq:Lboundrecur} requires that 
\be
g_3(z_1^2,z_2^2,z_3^2) = \chi_{(1,0,0)}(z_1^2,z_2^2,z_3^2).
\ee
We have verified that this indeed comprises the full solution for $L=3$.
Computing $\psi_{(()} = s_3 \psi_{(((}$, and setting $z_3=qz_2$, we find the recursion between size $L=3$ and size $L=1$:
\be
\left.\psi_{(()}\right|_{z_3=qz_2}=- \frac{A_3}{A_1}k(z_2,\zeta_1)^2k(z_2,\zeta_2)^2k(z_2,z_1)^4\psi_(.
\ee

Assuming that $p(z_i;z_1,\ldots,\hat z_i,\hat z_{i+1},\ldots,z_L)$ takes similar forms for each $L$, and noting that $p$ must be symmetric in all $z_j\neq z_i$, we find the general form to be:
\be
\begin{split}
L=2,\qquad\qquad\; p(z_i)&=-\frac{A_2}{A_0}k(z_i,\zeta_1)^2k(z_i,\zeta_2)^2,\\
L=3,\qquad\quad p(z_i;z_j)&=-\frac{A_3}{A_1}k(z_i,\zeta_1)^2k(z_i,\zeta_2)^2k(z_i,z_j)^4,\qquad j\neq i,i+1\\
p(z_i;z_1,\ldots,\hat z_i,\hat z_{i+1},\ldots,z_L)&=-\frac{A_L}{A_{L-2}}k(z_i,\zeta_1)^2k(z_i,\zeta_2)^2\prod_{j\neq i,i+1}k(z_i,z_j)^4.
\end{split}
\ee

Using similar arguments, it can be deduced that the proportionality factors $r_0$ and $r_L$ in the boundary recursions \eqref{eq:boundrecurpsi} are given by
\be
r_0(z_2,\ldots,z_L;\zeta_1) = (-1)^{L+1}\frac{A_L}{A_{L-1}} k(\zeta_1,\zeta_2) \prod_{i=2}^{L} k(\zeta_1,z_i)^2,
\ee
and
\be
r_L(z_1,\ldots,z_{L-1};\zeta_2) = (-1)^{L+1}(s^2)\frac{A_L}{A_{L-1}} k(1/s\zeta_2,s\zeta_1)\prod_{i=1}^{L-1}k(1/s\zeta_2,sz_i)^2.
\ee

\subsubsection{Recursion for components}
We would like to find recursions relating certain components of $\ket\Psi$ for size $L$ to components for size $L-1$ and $L-2$, as we already have done for small system sizes. Such a recursion would allow us to determine the still unkown functions $f_L$ and $\tilde f_L$ in \eqref{eq:psidef1} and \eqref{eq:psidef2}. Recalling the definition of $s_i$ in \eqref{sidef}, we have that 
\be
\psi_{(\ldots()}=s_L \psi_{(\ldots(}\ ,
\ee
and since $\left.\psi_{(\ldots(}\right|_{z_L=qz_{L-1}}=0$, it follows that
\begin{align}
\left.\psi^L_{(\ldots()}\right|_{z_L=qz_{L-1}} &=\left.\left(\pi_L\frac{k(sz_L,s\zeta_2)}{[q][s^2z_L^2]}\prod_{0\leq i<j\leq L}k(z_j,z_i)f_L(z_1,\ldots,z_L)\right)\right|_{z_L=qz_{L-1}}\non\\
&=k(z_{L-1},\zeta_2)\prod_{i=0}^{L-2}k(z_{L-1},z_i)^2\prod_{0\leq i<j\leq L-2}k(z_j,z_i)f_L(z_1,\ldots,z_{L-1},s^2qz_{L-1})\non\\
&=k(z_{L-1},\zeta_1)^2k(z_{L-1},\zeta_2)\prod_{i=1}^{L-2}k(z_{L-1},z_i)^2\frac{f_L(z_1,\ldots,z_{L-1},s^2qz_{L-1})}{f_{L-2}(z_1,\ldots,z_{L-2})}\psi^{L-2}_{(\cdots(}.
\end{align}
Here we have used the properties of $k$ given below \eqref{eq:L2recur}. From above, the proportionality factor in this relation is given by $p(z_{L-1};z_1,\ldots,z_{L-2})$, so we arrive at a recursion for $f_L$:
\be
f_L(z_1,\ldots,z_{L-1},s^2qz_{L-1})=-\frac{A_L}{A_{L-2}}k(z_{L-1},\zeta_2)\prod_{j=1}^{L-2} k(z_{L-1},z_j)^2f_{L-2}(z_1,\ldots,z_{L-2}).
\ee
A similar argument finds a recursion for $\tilde f_L^{\mbox{}}$. Due to the symmetry properties of both these functions, the recursions can be generalised to arbitrary $i$:
\be
\lb
{eq:recurSym}\begin{split}
f_L(z_1,\ldots,z_i,s^2qz_i,\ldots,z_L)&=-\frac{A_L}{A_{L-2}}k(z_i,\zeta_2)\prod_{j\neq i,i+1} k(z_i,z_j)^2f_{L-2}(z_1,\ldots,\hat z_i,\hat z_{i+1},\ldots,z_L),\\
\tilde f_L(z_1,\ldots,z_i,s^2qz_i,\ldots,z_L)&=-\frac{A_L}{A_{L-2}}k(z_i,\zeta_1)\prod_{j\neq i,i+1} k(z_i,z_j)^2\tilde f_{L-2}(z_1,\ldots,\hat z_i,\hat z_{i+1},\ldots,z_L).\\
\end{split}
\ee

The boundary recursion \eqref{eq:boundrecurpsi} can be immediately applied to the extremal components \eqref{eq:psidef1} and \eqref{eq:psidef2}, and we find that $f$ and $\tilde f$ in addition satisfy
\be
\lb
{eq:recurSymbound}
\begin{split}
f_{L}(z_1,\ldots,z_{L-1},\zeta_2/q;\zeta_1,\zeta_2) &=(-1)^{L+1}\frac{A_L}{A_{L-1}}\prod_{j=1}^{L-1}k(1/\zeta_2,z_j)\ f_{L-1}(z_1,\ldots,z_{L-1};\zeta_1,\zeta_2/q),\\
\tilde f_{L}(q\zeta_1,z_2,\ldots,z_L;\zeta_1,\zeta_2) &=(-1)^{L+1}s^2\frac{A_L}{A_{L-1}}\prod_{j=2}^L k(\zeta_1,z_j)\ \tilde f_{L-1}(z_2,\ldots,z_L;q\zeta_1,\zeta_2),
\end{split}
\ee
where we have explicitly indicated the dependences on $\zeta_1$ and $\zeta_2$.

\subsection{Degree}
Polynomial solutions of the $q$KZ can be labeled by their top degree $\mu$, where $\mu$ is a partition, $\mu_1\geq \mu_2 \geq \ldots \geq \mu_L\geq 0$. These solutions are of the form
\be
\sum_{\nu \in W\cdot\mu} c_{\nu} z^{2\nu},
\ee
where $W\cdot \mu$ denotes the orbit of $\mu$ under the action of the Weyl $W$ group of type BC$_L$, $c_\nu$ are constants, and
\[
z^{2\mu} =\prod_{i=1}^L z_i^{2\mu_i}.
\]

We can use the recursions \eqref{eq:recurSym} and \eqref{eq:recurSymbound} to find out what the minimal degree of $f_L$ has to be for arbitrary size. Consider $i=1$ and denote the top degree of $f_L$ by $\nu^{(L)}=(\nu_1^{(L)},\ldots,\nu_L^{(L)},0,0,\ldots)$. Since the degree of $k(z_1,z_j)$ is $(1,0)$ in the variables $z_1^2$ and $z_j^2$, the top degree in $z_1^2$ on the right hand side of \eqref{eq:recurSym} is $2L-3$. Comparing top degrees in \eqref{eq:recurSymbound}, it immediately follows that $\nu_{j}^{(L)}$ is at least equal to $\nu_{j}^{(L-1)}+1$. We thus find that the following inequalities have to hold,
\begin{align}
\nu_1^{(L)} + \nu_2^{(L)} \geq 2L-3,
\label{eq:degree1}\\
\nu_j^{(L)} \geq \nu_{j}^{(L-1)}+1. 
\end{align}
For a possible minimal degree solution these inequalites become equalities, and we find that
\be
\nu_j^{(L)} = L-j\qquad (j=1,\ldots,L),
\ee
which agrees with the solutions we explicitly constructed for the small system sizes $L=1,2,3$ in Section~\ref{se:smallsize}. We will write $\nu^{(L)}=\lambda^{(L)}+\lambda^{(L+1)}$, where $\lambda^{(L)}$ is the partition of $|\lambda^{(L)}|= \left\lceil \frac{L}{2}\left( \frac{L}{2}-1\right)\right\rceil$ with 
\be
\lb{eq:lambda}\lambda_j^{(L)} = \left\lfloor \frac{L-j}{2}\right\rfloor\qquad j=1,\ldots,L,
\ee
i.e.
\be
\lambda^{(2n)}=(n-1,n-1,\ldots,1,1,0,0),\qquad \lambda^{(2n+1)}=(n,n-1,n-1\ldots,1,1,0,0).
\ee

From the degree of $k(z_j,z_i)$ it immediately follows that the product of factors in the expression for the extremal components amount to a degree of $\lambda^{(L+1)}+\lambda^{(L+2)}$. Solutions of the $q$KZ equation of minimal degree, which are relevant for the O($n=1$) loop model with open boundaries, therefore have degree $\mu^{(L)}$, with
\be
\mu^{(L)} = \lambda^{(L)}+2\lambda^{(L+1)}+\lambda^{(L+2)},
\ee
so that
\be
\mu_j^{(L)} = 2L+1-2j.
\label{degree}
\ee
The total degree of these solutions is equal to $|\mu^{(L)}|=L^2$ and the degree in each variable $z_i^2$ is equal to $\mu_1 = 2L-1$.

\subsection{Solution}
For $L=2$ and $L=3$, the solution contains a symmetric function which involves the symplectic character defined in \eqref{sympchar}. The solution for general $L$ can also be expressed in terms of a special symplectic character. It turns out that the following two functions satisfy the necessary recursions \eqref{eq:recurSym} and \eqref{eq:recurSymbound}, and have the correct degree $\nu^{(L)}$,
\be
\begin{split}
f_L(z_1,\ldots,z_L)&=A_L\chi_{\lambda^{(L+1)}}(z_1^2,\ldots,z_L^2,\zeta_2^2)\chi_{\lambda^{(L)}}(z_1^2,\ldots,z_L^2)\\
\tilde f_L(sz_1,\ldots,sz_L) &=A_L(s^2)^L\chi_{\lambda^{(L+1)}}(s^2\zeta_1^2,s^2z_1^2,\ldots,s^2z_L^2)\chi_{\lambda^{(L)}}(s^2z_1^2,\ldots,s^2z_L^2).
\end{split}
\label{Symdef}
\ee
The classical character $\chi_\lambda$ for the partition $\lambda=\lambda^{(L)}$ appears repeatedly in related studies on loop models \cite{DF05,ZJ07} and symmetry classes of alternating sign matrices \cite{Okada06}. It is further worthwhile noting that \eqref{eq:recurSym} is satisfied because of the recursion
\be
\lb{eq:chirecur}
\chi_{\lambda^{(L)}}(z_1^2,\ldots,z_L^2)|_{z_{j+1}=qz_j}=(-1)^L\prod_{i\neq j,j+1}k(z_j,z_i)\ \chi_{\lambda^{(L-2)}}(z_1^2,\ldots,\hat z_j^2,\hat z_{j+1}^2,\ldots,z_L^2),
\ee
and the specification $s^4=1$.

\subsection{Normalisation}
\label{se:normalisation}
The normalisation of the groundstate eigenvector is defined by
\be
Z_L=\bra \Psi\Psi\rangle_L,
\ee
which can also be expressed as the sum over all the coefficients of $\ket\Psi_L$,
\be
Z_L=\sum_{\alpha \in {\rm LP}_L}\psi_{\alpha},
\label{eq:norm}
\ee
as the left eigenvector of the transfer matrix satisfies $\bra \Psi\alpha\rangle=1$ for all $\alpha$.

We have derived in \eqref{eq:Zrecur} and \eqref{eq:Zrecurbound} the recursions for the normalisation $Z_L$. Using the recursion \eqref{eq:chirecur} for the symplectic character $S_L(\ldots,z_i,\ldots)=\chi_{\lambda^{(L)}}(\ldots,z_i^2,\ldots)$ defined in \eqref{sympchar}, we thus note that
\begin{align}
S_{L+2}&(\zeta_1,z_1,\ldots,z_L,\zeta_2)S_{L+1}(\zeta_1,z_1,\ldots,z_L)S_{L+1}(z_1,\ldots,z_L,\zeta_2)S_L(z_1,\ldots,z_L)|_{z_{i+1}=qz_i}\non\\
&=k(z_i,\zeta_1)^2\ k(z_i,\zeta_2)^2\prod_{j\neq i,i+1} k(z_i,z_j)^4\ S_L(\zeta_1,\ldots,\hat z_i,\hat z_{i+1},\ldots,\zeta_2)\non\\
&\quad\times S_{L-1}(\zeta_1,\ldots,\hat z_i,\hat z_{i+1},\ldots)S_{L-1}(\ldots,\hat z_i,\hat z_{i+1},\ldots,\zeta_2)S_{L-2}(\ldots,\hat z_i,\hat z_{i+1},\ldots).
\end{align}
Since the recursions \eqref{eq:Zrecur} and \eqref{eq:Zrecurbound} specify enough points to uniquely determine $Z_L$ of degree $\mu^{(L)}=\lambda^{(L)}+2\lambda^{(L+1)}+\lambda^{(L+2)}$, see \eqref{degree}, and, when we set $A_L=(-1)^LA_{L-1}$, this product of four symplectic characters satisfies the same recursions, we conclude that
\begin{align}
Z_L(z_1,\ldots,z_L)&=S_{L+2}(\zeta_1,z_1,\ldots,z_L,\zeta_2)S_{L+1}(\zeta_1,z_1,\ldots,z_L)\non\\
&\quad\times S_{L+1}(z_1,\ldots,z_L,\zeta_2)S_L(z_1,\ldots,z_L).
\end{align}
This proof also requires equivalence for $L=1$ and $L=2$, which is easy to show. In particular, the normalisation of the groundstate of the Hamiltonian \eqref{eq:ham} is obtained by setting $z_i=1$, and is given by
\be
Z_L = \tilde{Z}_2(c_1,c_2)\tilde{Z}_{1}(c_1)\tilde{Z}_{1}(c_2)\tilde Z_0,
\label{eq:homonorm}
\ee
where
\be
\begin{split}
\tilde{Z}_{0} &= S_L(1,\ldots,1)\\
\tilde{Z}_{1}(c_i) &= S_{L+1}(\zeta_i,1,\ldots,1),\\
\tilde{Z}_{2}(c_1,c_2) &= S_{L+2}(\zeta_1,1,\ldots,1,\zeta_2),
\end{split}
\ee
and
\be
c_i = \frac{3}{1+\zeta_i^2+\zeta_i^{-2}}\ .
\ee

\section{Conclusion}
We have given an explicit description, for \emph{finite} system sizes and without resorting to the Bethe Ansatz, of the groundstate of the O($n=1$) loop model with open boundaries. The boundary conditions considered in this paper contains as special cases those of reflecting and mixed boundary conditions which have been considered before \cite{DF05,ZJ07,GierP07}. In an alternative interpretation the O($n=1$) is equivalent to the stochastic raise and peel model \cite{PearceRGN02,Pyatov04} for which the groundstate is a stationary state distribution. In this setting it is important to compute the normalisation so that the stationary state is a properly normalised probability distribution. The derivation of the normalisation of the raise and peel stationary state, or O($n=1$) groundstate, with two open boundaries is presented in Section~\ref{se:normalisation}. 

There is another, independent reason for computing the normalisation, which is in the context of the Razumov-Stroganov conjecture \cite{BGN01,RS01,Gier02}. This conjecture states that there is an intriguing relation between the O($n=1$) groundstate $\ket\Psi_L$ and the combinatorics of a fully packed loop (FPL) model on finite geometries, as well as other combinatorial objects such as alternating sign matrices and symmetric plane partitions~\cite{Bressoud}. In particular it states that the groundstate normalisation is equal to the statistical mechanical partition function of an FPL model on a certain finite patch of the square lattice. Such a link has been made for the groundstate of the model with \textit{identified} open boundaries, see \cite{GierR04}, but for the model considered here, which genuinely has two open boundaries, it is not known which FPL geometry gives rise to a partition function equal to the normalisation of $\ket\Psi_L$ as computed in \eqref{eq:homonorm}. Understanding the underlying combinatorics for the general case of two boundaries will therefore lead to a deepening of our understanding of the RS conjecture, as well as to possible generalisations of symmetric plane partitions and alternating sign matrices.

We hope and expect that our results will lead to explicit expressions for finite size correlation functions of the open O($n=1$) loop model, as well as for those of the closely related XXZ quantum spin chain with anisotropy $\Delta=-1/2$ and non-diagonal open boundaries at both ends. Other incarnations of the model considered here to which our results may be applied include the conformally invariant stochastic raise and peel model \cite{PearceRGN02,Pyatov04}, and supersymmetric lattice models with boundaries \cite{YangF04,GNPR05}.

One way forward would be to express certain linear combinations of the components of the grounstate eigenvector in terms of multiple contour integrals such as was done for reflecting boundary conditions \cite{DFZJ07,DFZJ07b}. However, we anticipate some fundamental difficulties with this approach for the case of open boundaries, related also to the lack of convenient factorised expressions which do exist for reflecting and mixed boundary conditions \cite{GierP07}. We further hope to make a connection between our solutions and Macdonald-Koornwinder polynomials of type $(C_n^\vee,C_n)$ for specialised parameters \cite{Kasatani}, as well as with those in the form of Jackson integrals for $q$KZ equations on tensor product spaces, see \cite{VarT,Mim} for the case of type A.

\section*{Acknowledgments}
JdG and AP acknowledge financial support from the Australian Research Council. KS is supported by the ANR Research Project``Boundary integrable models: algebraic structures and correlation functions'', contract number JC05-52749. We are grateful for the kind hospitality offered by the ESI in Vienna where part of this work was undertaken. We furthermore warmly thank Pascal Baseilhac, Pavel Pyatov and Arun Ram for many stimulating discussions and encouragements, and in particular Paul Zinn-Justin and Luigi Cantini for sharing with us their solution for $L=2$ and for pointing out an omission in an earlier version of this paper. 
\appendix
\section{Proof of the recursion \eqref{eq:Tphi}}
\label{ap:recursion}
In the following we will use the shorthand notation 
\[
a(z)=\frac{[q/z]}{[qz]},\qquad b(z)=-\frac{[z]}{[qz]}.
\] 
We first define $\ket{\alpha'}_L=\varphi_i\ket{\alpha}_{L-2}$ to be the link pattern of length $L$ with a small link connecting sites $i$ and $i+1$ inserted into the link pattern $\ket{\alpha}_{L-2}$. Restricting our focus to the action of the transfer matrix on the sites $i$ and $i+1$, we find
\be
T_L(w;z_{i+1}=qz_i)\ket{\alpha'}_L=\quad\raisebox{-40pt}{\includegraphics[height=100pt]{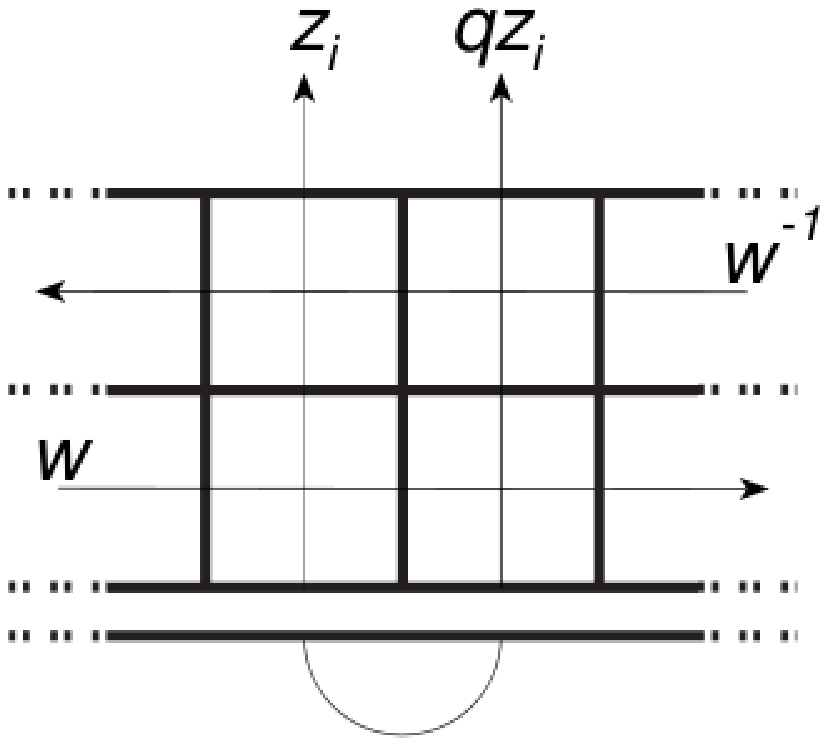}},
\ee
As each R-operator consists of two terms, the action of $T_L$ on sites $i$ and $i+1$ produces sixteen terms,
\begin{align}
&a(z_iw)a(qz_iw)a(w/z_i)a(w/qz_i)\quad\raisebox{-22pt}{\includegraphics[height=50pt]{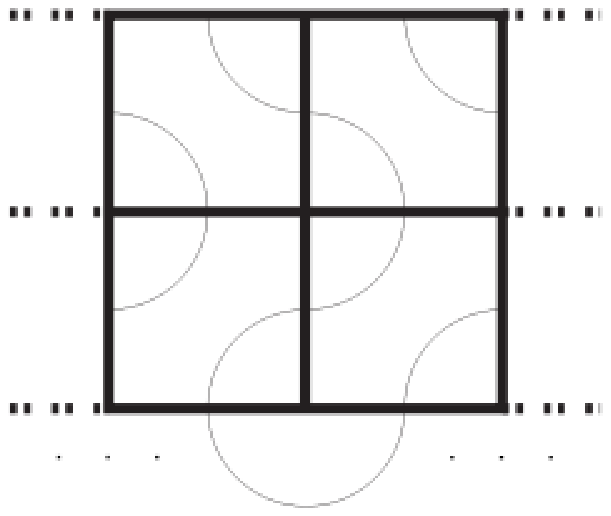}}\,+\non\\
&a(z_iw)a(qz_iw)a(w/z_i)b(w/qz_i)\quad\raisebox{-22pt}{\includegraphics[height=50pt]{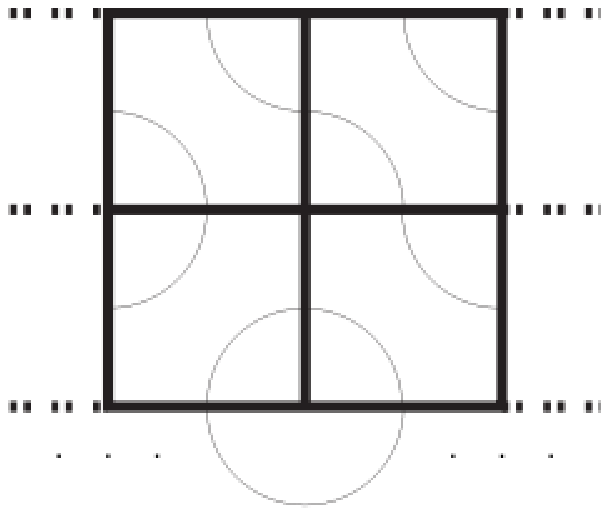}}\,+\,\ldots\,.
\end{align}
Some of these pictures are equivalent with respect to their external connectivities. In total there are five different kinds of connectivities. For instance, one of the connectivities has
\begin{align}
&a(z_iw)b(qz_iw)a(w/z_i)b(w/qz_i)\quad\raisebox{-22pt}{\includegraphics[height=50pt]{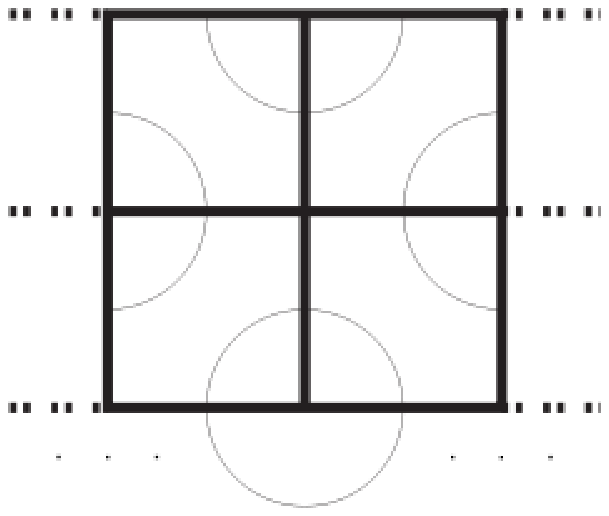}}\,+\non\\
&a(z_iw)b(qz_iw)b(w/z_i)b(w/qz_i)\quad\raisebox{-22pt}{\includegraphics[height=50pt]{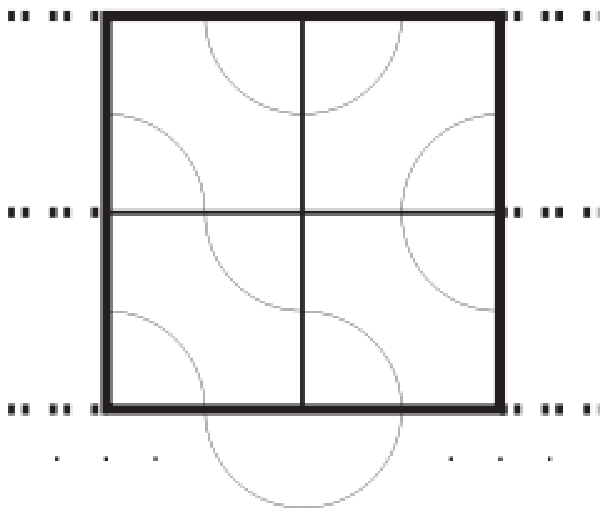}}\,+\non\\
&a(z_iw)b(qz_iw)a(w/z_i)a(w/qz_i)\quad\raisebox{-22pt}{\includegraphics[height=50pt]{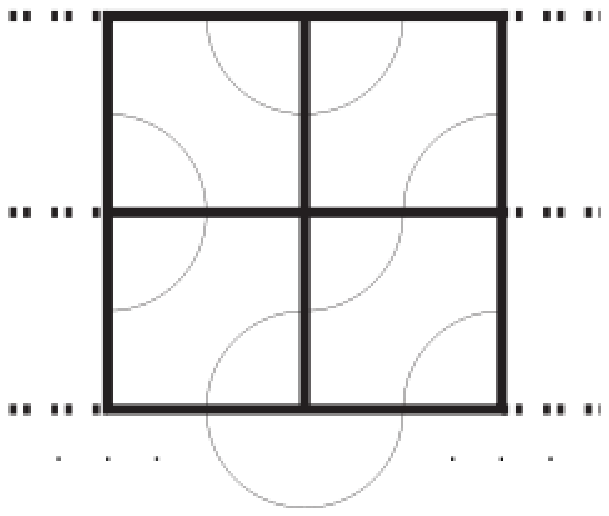}}.
\end{align}
The closed loop in the first diagram is erased at the expense of a factor $-(q+q^{-1})$, after which the coefficients of the three diagrams sum to $0$. Using the fact that $a(qu)a(u)+b(qu)b(u)-{(q+q^{-1})}a(qu)b(u)=0$, it is easy to show that this happens for three of the remaining four kinds of connectivities as well, and we are left with
\begin{align}
&T_L(w;z_{i+1}=qz_i)\ket{\alpha'}_L\non\\
&\qquad=a(z_iw)b(qz_iw)b(w/z_i)a(w/qz_i)\quad\raisebox{-38pt}{\includegraphics[height=80pt]{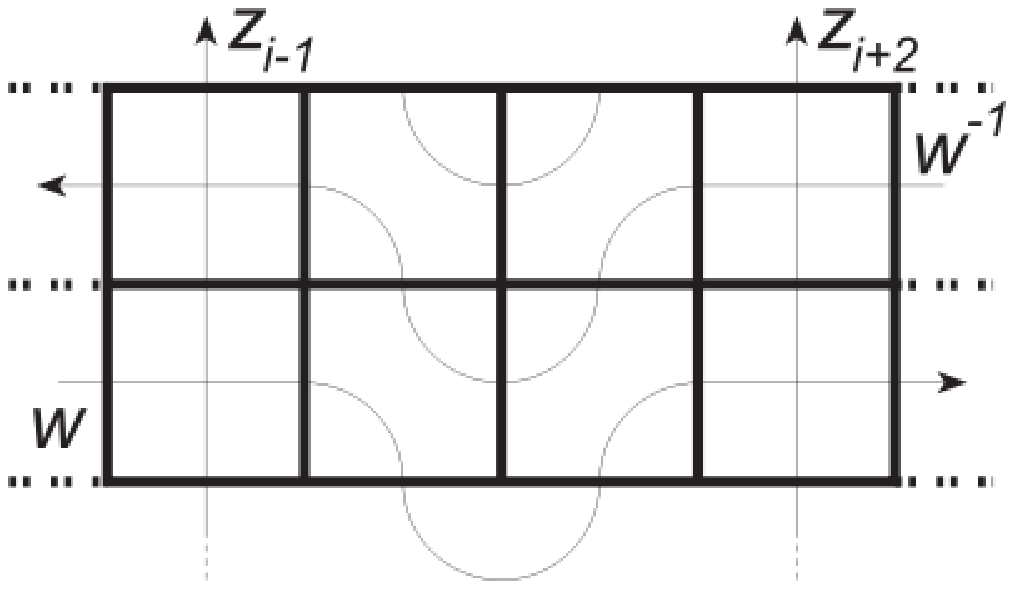}}\non\\
&\qquad=a(z_iw)b(qz_iw)b(w/z_i)a(w/qz_i)\ \varphi_i\quad\raisebox{-38pt}{\includegraphics[height=80pt]{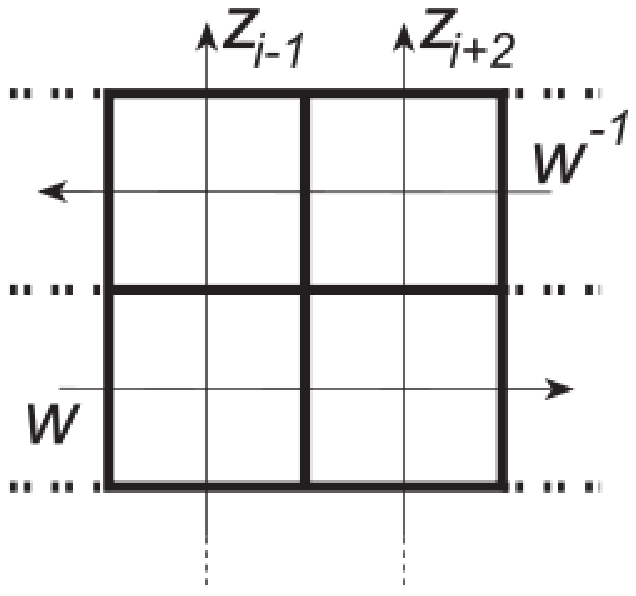}}\non\\
&\qquad=\frac{[q/z_iw][q^2z_i/w]}{[q^2z_iw][qw/z_i]}\varphi_i\, T_{L-2}(w;\hat z_i,\hat z_{i+1})\ket{\alpha}_{L-2}.
\end{align}

\section{Example: $L=3$}
\label{ap:L=3}
For this example we will take $q=\e^{2\pi \ii/3}$ and $s^4=1$, but we will leave $b$ generic. For $N=3$, we use the notation
\begin{align*}
\psi_1&=\psi_{(((}\hspace{3cm}\psi_5=\psi_{)((}\\
\psi_2&=\psi_{(()}\hspace{3cm}\psi_6=\psi_{)()}\\
\psi_3&=\psi_{()(}\hspace{3cm}\psi_7=\psi_{))(}\\
\psi_4&=\psi_{())}\hspace{3cm}\psi_8=\psi_{)))},
\end{align*}
and we recall the definition \eqref{sidef},
\be
s_i = q+q^{-1} -a_i =-1-a_i.
\ee

Considering in turn each $i$ and $\alpha$, the $q$KZ equation
\be
\sum_\alpha \psi_\alpha \left(e_i\ket\alpha\right)=-\sum_\alpha \left(a_i\psi_\alpha\right) \ket\alpha
\ee
implies the following $32$ system equations:
\begin{equation}
\left.\begin{array}{r@{\;}c@{\;}l}
a_0\psi_\alpha & = & 0 \\
s_0\psi_{\alpha+4} &= & \psi_\alpha
\end{array}
\right\}
\qquad \alpha=1,\ldots,4
\end{equation}

\begin{align}
a_1\psi_\alpha &=0 \qquad\qquad\qquad\quad \alpha=1,2,5,\ldots,8\non\\
s_1\psi_3&=\psi_1+\psi_2+b\psi_5+\psi_7\label{s1}\\
s_1\psi_4&=\psi_6+\psi_8\non
\end{align}

\begin{align}
a_2\psi_\alpha &=0 \qquad\qquad\qquad\quad \alpha=1,3,4,5,7,8\non\\
s_2\psi_6&=\psi_4+\psi_5+\psi_7+\psi_8\label{s2}\\
s_2\psi_2&=\psi_1+\psi_3\non
\end{align}

\begin{equation}
\begin{array}{r@{\;}c@{\;}ll}
a_3\psi_{2\alpha} &=& 0\qquad \qquad & \alpha=1,\ldots,4 \\
s_3\psi_{2\alpha-1} &=& \psi_{2\alpha} & \alpha=1,3\\
s_3\psi_{2\alpha-1} &=& b\psi_{2\alpha} & \alpha=2,4
\end{array}
\end{equation}

The relations where the action of the projector $a_i$ gives zero force certain symmetry restrictions on the components. For instance,
\be
\label{eq:psisym}
\begin{split}
\psi_1&=k(z_1,\zeta_1)k(z_2,z_1)k(z_2,\zeta_1)k(z_3,z_2)k(z_3,z_1)k(z_3,\zeta_1)\,f_1(z_1,z_2,z_3),\\
\psi_8&=k(1/sz_1,1/sz_2)k(1/sz_2,1/sz_3)k(1/sz_1,1/sz_3)k(1/sz_3,1/s\zeta_2)\\
&\qquad\times k(1/sz_2,1/s\zeta_2)k(1/sz_1,1/s\zeta_2)\,f_8(sz_1,sz_2,sz_3),
\end{split}
\ee
where $f_\alpha(z_1,z_2,z_3)$ is a symmetric function invariant under $z_i \rightarrow 1/z_i$.

Using the system equations we may obtain the components $\psi_\alpha$ in terms of $\psi_1$ and $\psi_8$. Assuming we know $\psi_1$, we find $\psi_2=s_3\psi_1$ , then $\psi_3=s_2\psi_2-\psi_1$, and $\psi_4=b^{-1} s_3\psi_3$. Then we can find $\psi_8$ from
\begin{align}
\psi_8&=s_2\psi_6-\psi_4-\psi_5-\psi_7,
\label{psi8}
\end{align}
by using $\psi_6=s_1\psi_4-\psi_8$ and applying $s_3$ on both sides of \eqref{psi8} to get
\begin{align}
-\psi_8&=s_3s_2(s_1\psi_4-\psi_8)+\psi_4-\psi_6-b\psi_8,
\end{align}
which implies
\begin{align}
(b-1)\psi_8&=(s_3s_2s_1-s_1+1)\psi_4.
\end{align}
The expressions for $\psi_2$, $\psi_3$, $\psi_4$ as well as $\psi_8$ can be neatly rewritten in a factorised form as in \cite{GierP07}. However, if $b=1$, the component $\psi_8$ cannot be determined this way. 

In a similar way, given $\psi_8$ we can find $\psi_4$, then $\psi_6$, and $\psi_2$. Since we can express $\psi_4$ and $\psi_2$ in two different ways, these have to satisfy certain consistency conditions. Now we can find the remaining two components,
\be
\begin{split}
(b-1)\psi_5&=s_1\psi_3-\psi_1-\psi_2-s_2\psi_6+\psi_4+\psi_8\\
\psi_7&=s_1\psi_3-\psi_1-\psi_2-b\psi_5.
\end{split}
\ee
Again, if $b=1$ these two components cannot be found separately in this way. However, their sum can be determined. Assuming we find an expression for one of these components (say, by solving $s_0\psi_5=\psi_1$ for $\psi_5$) which satisfies the appropriate degree and the symmetries imposed by \eqref{s1} and \eqref{s2}, the entire system can be shown to be consistent.

\section{Symmetry of proportionality factor}
\lb{ap:propsym}

Denoting the proportionality factor of \eqref{eq:psiprop} as $p_i(z_i;z_1,\ldots,\hat z_i,\hat z_{i+1},\ldots,z_L)$, we want to show that $p_i$ takes the same form for each $i$. To do this we consider the normalisation
\be
Z_L=\bra \Psi\Psi\rangle_L,
\ee
which can also be written as the sum over the components of $\ket\Psi$, as in \eqref{eq:norm}. Acting with $\bra\Psi$ on both sides of the $q$KZ equation \eqref{qKZTL_TypeC2}, and using that $\pi_i$ commutes with $\bra\Psi$, we have
\begin{align}
\pi_iZ_L&=\bra\Psi\check R_i(z_i/z_{i+1})\ket\Psi_L\non\\
&=\sum_\alpha\psi_{L,\alpha} \bra\Psi \check R_i(z_i/z_{i+1})\ket\alpha\non\\
&=\sum_\alpha\psi_{L,\alpha}\left(\frac{[qz_{i+1}/z_i]}{[qz_i/z_{i+1}]} \bra\Psi\alpha\rangle -\frac{[z_i/z_{i+1}]}{[qz_i/z_{i+1}]}\bra\Psi e_i\ket\alpha\right).
\end{align}
Since $e_i\ket\alpha=\ket{\alpha'}$ for some link pattern $\alpha'$, and $\bra \Psi\alpha\rangle=1$ for all $\alpha$, this becomes
\begin{align}
\pi_iZ_L&=\sum_\alpha\psi_{L,\alpha}\left(\frac{[qz_{i+1}/z_i]}{[qz_i/z_{i+1}]}-\frac{[z_i/z_{i+1}]}{[qz_i/z_{i+1}]}\right)\non\\
&=\sum_\alpha\psi_{L,\alpha}=Z_L.
\end{align}
Similar arguments can be made to show that $\pi_0 Z_L=Z_L$ and $\pi_L Z_L=Z_L$. We therefore know that $Z_L$ remains unchanged under any permutation of the variables $z_i$. Recalling that
\be
\sum_\alpha\psi_{L,\varphi_i\alpha}(z_{i+1}=qz_i)\ \varphi_i\ket\alpha=p_i(z_i;\ldots,\hat z_i,\hat z_{i+1},\ldots)\sum_\alpha\psi_{L-2,\alpha}(\hat z_i,\hat z_{i+1})\ \varphi_i\ket\alpha,
\ee
we have
\be
Z_L(z_{i+1}=qz_i)=p_i(z_i;\ldots,\hat z_i,\hat z_{i+1},\ldots)Z_{L-2}(\hat z_i,\hat z_{i+1}).
\label{eq:Zrecur}
\ee
Since taking $z_i\rightarrow z_j$ would give the same result, we know that $p_i=p_j$ for all $i$ and $j$. We henceforth drop the index $i$ from $p_i$.

Completely analogously we can derive boundary recursions for the normalisation, which result in
\be
\begin{split}
Z_L(z_1=q\zeta_1,\ldots,z_L;\zeta_1,\zeta_2) &= r_0(z_2,\ldots,z_{L};\zeta_1) Z_{L-1}(z_2,\ldots,z_{L};q\zeta_1,\zeta_2),\\
Z_L(z_1,\ldots,z_L=\zeta_2/q;\zeta_1,\zeta_2) &= r_L(z_1,\ldots,z_{L-1};\zeta_2) Z_{L-1}(z_1,\ldots,z_{L-1};\zeta_1,\zeta_2/q).
\label{eq:Zrecurbound}
\end{split}
\ee

\newpage
\newcommand\arxiv[1]{\href{http://arxiv.org/pdf/#1}{\texttt{arXiv:#1}}}

\end{document}